\documentclass{article}
\usepackage[utf8]{inputenc}
\usepackage{amsfonts}
\usepackage{spconf,amsmath,graphicx}
\usepackage{xcolor}
\graphicspath{{./}}
\usepackage{subfigure}
\usepackage[T1]{fontenc}
\usepackage{csquotes}
\usepackage{float}
\usepackage{bm}

\name{Qing Zou$^*$, Luis A. Torres$\dag$, Sean B. Fain$^*$, Mathews Jacob$^*$.}
\address{$^*$The University of Iowa, Iowa City. $\dag$ University of Wisconsin at Madison}
\begin{document}

\title{Dynamic imaging using motion-compensated smoothness regularization on manifolds (MoCo-SToRM)}

\maketitle

\begin{abstract}
We introduce an unsupervised deep manifold learning algorithm for motion-compensated dynamic MRI. We assume that the motion fields in a free-breathing lung MRI dataset live on a manifold. The motion field at each time instant is modeled as the output of a deep generative model, driven by low-dimensional time-varying latent vectors that capture the temporal variability. The images at each time instant are modeled as the deformed version of an image template using the above motion fields. The template, the parameters of the deep generator, and the latent vectors are learned from the k-t space data in an unsupervised fashion. The manifold motion model serves as a regularizer, making the joint estimation of the motion fields and images from few radial spokes/frame well-posed. The utility of the algorithm is demonstrated in the context of motion-compensated high-resolution lung MRI.
 \end{abstract}

\section{Introduction}
The non-ionizing nature of radiation in MRI often makes it an attractive option for patient groups that require serial follow-up. Unfortunately, the slow nature of MRI often makes it challenging to image moving organs (e.g., lungs) at high spatial resolution. While respiratory motion can be frozen by breath-holds, the maximum achievable spatial resolution is restricted by the breath-hold duration. In addition, many subjects with compromised pulmonary function often find it difficult to comply with long breath-holds. In recent years, several motion-resolved schemes using radial acquisition strategies have been introduced to minimize the problem. Unlike gating methods that only reconstruct one motion phase, motion-resolved schemes such as XD-GRASP \cite{feng2016xd} use the information from the central k-space samples to bin the collected data into several phases, followed by image recovery. Manifold approaches follow a conceptually similar approach, except that the data is not explicitly binned; the manifold structure of images is exploited using a kernel-based formulation \cite{nakarmi2017kernel,ahmed2020free}. Recently, unsupervised deep generative models that exploit the manifold structure of images have been shown to outperform the classical manifold methods \cite{zou2021dynamic,yoo2021time}. 

Motion-compensated reconstructions are a more data-efficient alternative to the above motion-resolved approaches when there are no dynamic changes in image contrast during the acquisition. For instance, \cite{huttinga2021nonrigid} shows that motion fields can be reliable when estimated from a few k-space samples when a high-resolution reference is available. Motion-compensated reconstruction strategies such as \cite{zhu2020iterative} perform XD-GRASP reconstruction of low-resolution images, followed by the registration of the image volumes to estimate the deformation fields. The different phases can be registered together and averaged to obtain a motion-compensated (MoCo) volume. Alternatively, the recovery of a volume from undersampled k-t space data using the above motion estimates can be posed as a non-linear compressed sensing (iMoCo) problem \cite{zhu2020iterative}. A challenge with these schemes is the complexity of the multi-step algorithm and its dependence on the accuracy of the estimates in each of the steps, including self-gating signals and motion estimates. 


The main focus of this work is to introduce a novel unsupervised manifold-based deep-learning motion-compensated reconstruction scheme. This approach can be seen as the generalization of the motion-resolved generative manifold methods \cite{zou2021dynamic,yoo2021time} to the motion-compensated setting. In particular, we assume that the motion fields live on a low-dimensional manifold parameterized by low-dimensional latent vectors. We model the motion fields as outputs of a deep generator, which is driven by the latent vectors (See Fig. 1 (b)). We assume that each image frame is a deformed version of the template image using the learned motion estimate and a spatial transformation or interpolation layer. The parameters of the generator and the latent vectors are jointly learned, such that the multichannel Fourier measurements of the images closely match the actual measurements. Because the motion fields are often much less complex than the images themselves, this approach is expected to be less data-demanding than motion-resolved approaches. We demonstrate the utility of the proposed scheme in high-resolution structural lung MRI. 

\section{Methods}

\begin{figure*}[!t]
\centering
           \subfigure[Motion-resolved rec. (XD-GRASP)]{\includegraphics[width=0.34\textwidth]{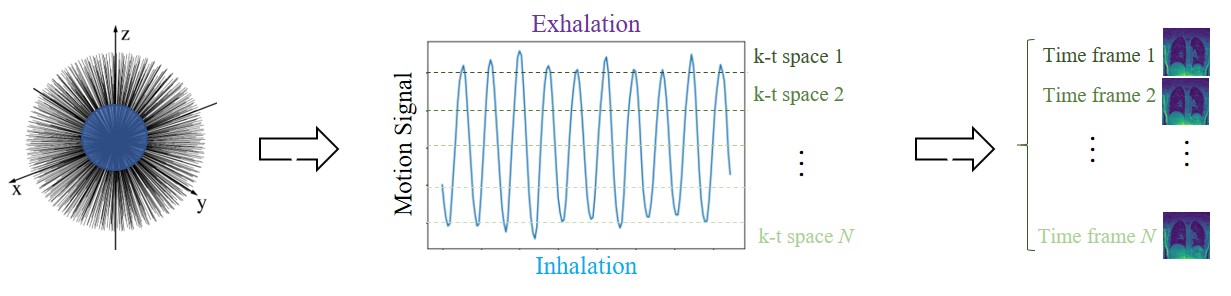}}\hspace{4em}
	\subfigure[Proposed MoCo-SToRM]{\includegraphics[width=0.56\textwidth]{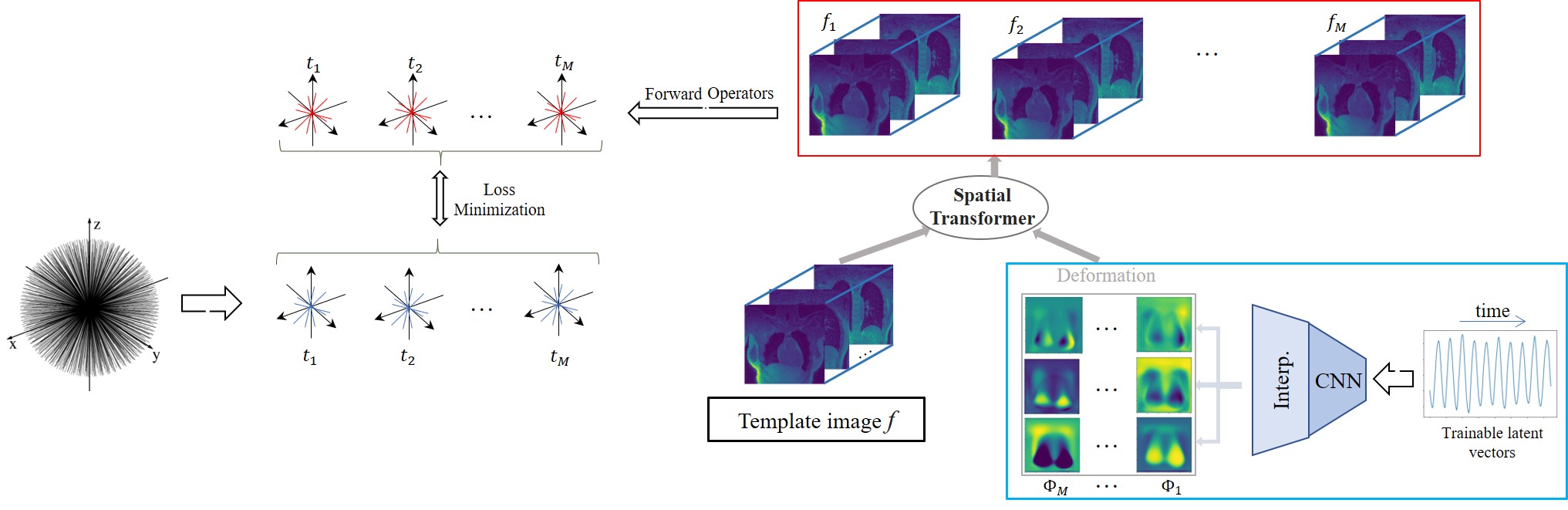}}
	\caption{\small{Illustration of XD-GRASP and proposed motion-compensated reconstruction algorithm. (a) XD-GRASP bins the k-t space data into different respiratory phases, followed by the joint recovery of the images using total variation and wavelet regularization. The respiratory signal is estimated from the central k-space data using coil clustering and low-pass filtering. 	(b) The proposed motion-compensated reconstruction jointly learns the motion vectors $\boldsymbol{\phi_t}$ and the static image template $f$ from the k-t space data. To regularize the motion fields, we model $\phi_t = \mathcal G_{\theta}(z(t))$ as the outputs of a deep CNN generator $\mathcal G_{\theta}$ whose weights are denoted by $\theta$, driven by low-dimensional (e.g., 1-D in lung MRI) latent vectors. The parameters of the CNN generator $\theta$, the latent vectors $z_t$, and the template $f$ are jointly estimated from the data. The loss is the mean-square error between the actual measurements and the multichannel measurements of the deformed images $f_t$, specified by \eqref{recon}.}}
	\label{illu}
\end{figure*}

\subsection{Proposed approach}
In this work, we generalize the motion-resolved deep manifold methods \cite{zou2021dynamic} to the motion-compensated reconstruction setting. We assume that the image volumes in the time series $f_t$ are deformed versions of a single image template $f$:
\begin{equation}\label{recon}
f_t(x,y,z) = f\Big(x-\phi_{x}(t),y-\phi_{y}(t),z-\phi_{z}(t)\Big) =  \mathcal{D}\left({f},\boldsymbol\phi(t)\right).
\end{equation}
Here, $\boldsymbol \phi(t) = \{\phi_{x}(t), \phi_{y}(t), \phi_{z}(t)\}$ is the motion field at the time instant $t$.
 The slow nature of MRI makes it impossible to fully acquire all the k-space samples of each time frame. This makes it difficult to recover the images, followed by the estimation of motion from them. 
 
 We propose to jointly estimate the motion field for each image and the template $f$ from the k-t space data. We note that the joint estimation problem is highly ill-posed. To regularize the motion fields, we use a manifold assumption. In particular, we assume that the motion field for each time frame $t$ is the output of a generative model 
 \begin{equation}
     \boldsymbol \phi(t)= \mathcal{G}_{\theta}(\mathbf{z}(t)),
 \end{equation}
 driven by a low-dimensional latent vector $\mathbf z(t) \in \mathbb R^d$, which captures the dynamics (e.g., respiratory motion in lung imaging). We set $d=1$ in our experiments in this paper. Thus, each image in the time series is expressed as the deformed version of the 3-D volume $f$, deformed by the motion field $\boldsymbol{\phi}_t$:
\begin{equation}\label{recon}
f_t(x,y,z) = \mathcal{D}\left({f},\underbrace{\mathcal{G}_{\theta}(\mathbf{z}_t)}_{\boldsymbol\phi(t)}\right).
\end{equation}
We implement $\mathcal{D}$ as an interpolation layer. With the above model, the reconstruction scheme amounts to the joint estimation of the image $f$, the parameters of the deep generator $\theta$, and the low-dimensional latent vectors $\mathbf Z = [\mathbf z_1,..,\mathbf z_M]$ from the measurements. We pose the reconstruction as the minimization of the cost function:
\begin{equation}\label{cost}
\mathcal{C}(\mathbf{z},\theta,{f}) = \sum_{t=1}^M||\mathcal{A}_t({f}_t)-\mathbf{b}_t||^2 + \lambda_t||\nabla_t\mathbf{Z}_t||,
\end{equation}
where ${f}_t$ is given by \eqref{recon}. Here, $\mathcal{A}_t$ are the forward models for each of the phases and $\mathbf b_t$ are the corresponding measurements. Note that we also add a smoothness penalty on the latent vector $\mathbf{z}$ along the time direction to encourage the latent vectors to be smooth. We implement the above networks in Pytorch and use Adam optimization to determine the optimal parameters $\mathbf{z},\theta$ and $f$. The network parameters are initialized as random, and the latent vectors are initialized as zero vectors. The image template $f$ is also initialized as zeros. 


The direct optimization of \eqref{cost} is associated with high computational complexity, especially for high-resolution images. To minimize the computational complexity, we propose a progressive training approach. Specifically, we solve \eqref{cost} for image volumes two times smaller than the original size in each dimension with the central k-space samples. The latent vectors and the network parameters that are learned from the lower resolution are used as initialization for the high-resolution setting. Because the motion fields are smooth, we use the same motion network for all resolutions; the interpolation layer is modified to derive the motion fields at the finer resolution. The static image $f$ is solved at the higher resolution.

\begin{figure}[!b]
	\centering
	\subfigure[Estimated Latent vectors]{\includegraphics[width=0.4\textwidth]{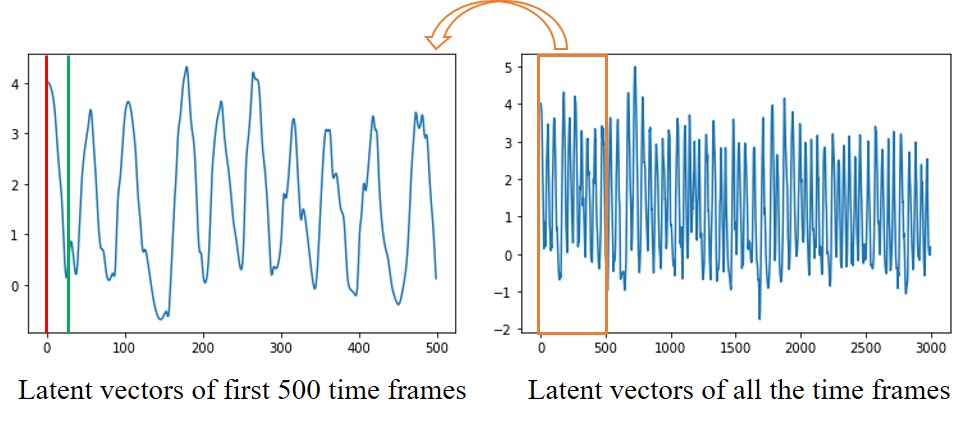}}\\
	\subfigure[Estimated flow maps]{\includegraphics[width=0.45\textwidth]{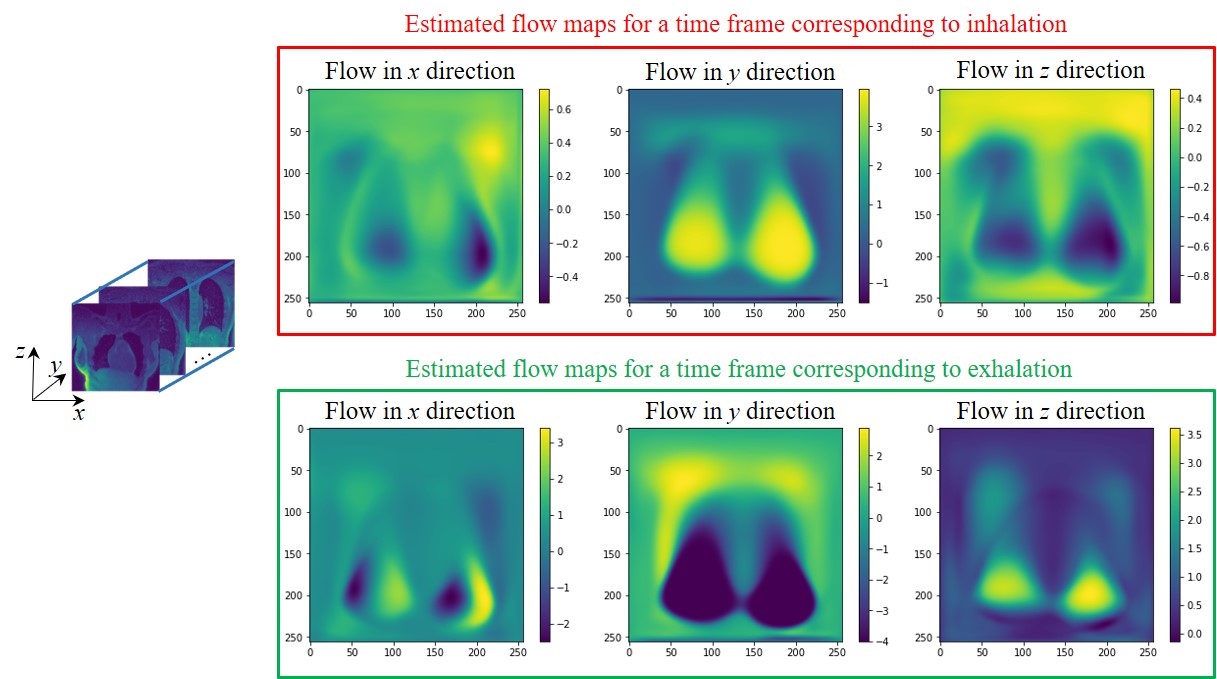}}\\\vspace{-1em}
	\caption{\small{Latent vectors and motion maps estimated using the proposed algorithm. (a) Latent vectors of the entire dataset (right) and a zoom of the first 500 frames (left). (b) Estimated flow maps in the inhalation and exhalation phases, marked by red and green lines in the latent vector plots, respectively.}}
	\label{lat}
\end{figure}

\section{Experiments}

\subsection{Datasets and imaging experiments}

The data used in the experiments in this work was acquired using an optimized 3D UTE sequence with variable-density readouts and a center of k-space oversampling \cite{johnson2013optimized} on a 1.5T GE scanner with no contrast administration. The variable-density readouts help retain SNR, and oversampling reduces aliasing artifacts. A bit-reversed ordering was used during the data acquisition. The data was acquired from an adult healthy subject using 8 coils. The prescribed FOV $= 32 \times 32\times 32$ cm$^3$. The matrix size is $256\times 256\times 256$. The data was acquired with 90K radial spokes with TR$\approx$ 3.2 ms and 655 samples/readout, corresponding to an approximately five-minute acquisition. Experimental results were generated using an Intel Xeon CPU \@ 2.4 GHz with an NVIDIA Titan V (CEO Edition) 32GB GPU.

\begin{figure*}[!h]
	\centering
	\includegraphics[height=0.45\textheight]{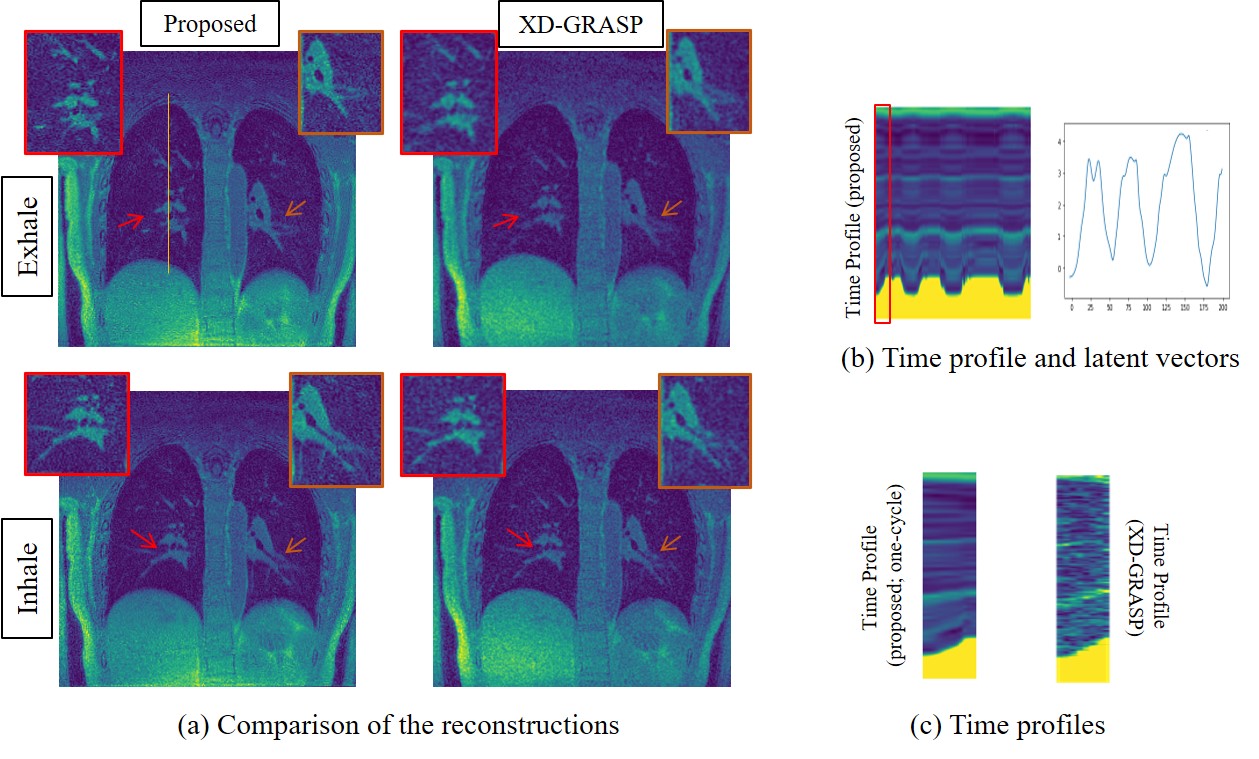}
	\caption{\small{Comparison of the proposed scheme and the optimized XD-GRASP approach. (a) Comparisons of the reconstructions from the two approaches. (b) Time profiles for the 200 time frames along the yellow line indicated in (a). The corresponding flipped latent vectors for the 200 time frames is also shown. (c) Comparisons for the time profiles of one cycle from the proposed approach and XD-GRASP. }}
	\label{comp}
\end{figure*}

\subsection{MoCo-SToRM}
We split the k-t space dataset into 3000 time frames, each consisting of 30 spokes/frame ($\approx$ 0.1 seconds/frame) for the proposed approach. Note that the recovery of each 3D time frame from 30 spokes is highly ill-posed. The CNN generator is fed with 1-D latent vectors $z_t$ to synthesize motion vectors $\boldsymbol{\phi}$ of a $64\times 64\times 64$ grid. The motion vectors are resampled to the final image grid and are used to deform the image template $f$. We use a two-step strategy, where $f$ is first solved on a $128\times 128\times 128$ grid, together with the associated motion fields first learned from the data. Once the algorithm converges, we solve the problem at the finest grid ($256\times 256\times 256$). The interpolation layer is modified to use the same generator and latent vectors at the finest resolution. 

The latent vectors and motion fields in two time frames learned by the algorithm are shown in Fig. \ref{lat}. We note that the latent vectors resemble respiratory motion with approximately 6 seconds/cycle on average. The motion fields corresponding to two time frames marked by red and green lines in Fig. \ref{lat}.(a) are shown in (b). We note that the proposed approach is able to generate detailed flow-maps, even though each time frame is sampled only with 30 spokes. The corresponding reconstructed images of the two time frames are shown in Fig. 3 (a).

\subsection{Comparison with the state of the art}

We compare the proposed unsupervised motion-compensated reconstruction with the motion-resolved reconstruction method in \cite{zhu2020iterative}, which is an optimized version of the XD-GRASP reconstruction \cite{feng2016xd}. The reconstruction results and comparisons are shown in Fig. \ref{comp}. The proposed approach recovered images with a temporal resolution of 0.1 seconds, while XD-GRASP recovered a few phases. We manually selected an inhalation phase and an exhalation state from the time series that best matched the corresponding XD-GRASP images. We also show some zoomed regions of the lung. From the comparisons, we see that the proposed motion-compensated reconstructions are less blurry and more well-defined compared to the optimized XD-GRASP reconstructions.The time profile of the proposed sections are shown in (b), while a zoomed sub-region indicated by the red rectangle is shown below to compare with the XD-GRASP time profiles. The variability in the respiratory cycles can be appreciated from the proposed cycles, which closely match the learned latent vectors. We also observe that the proposed approach offers smoother time profiles compared to XD-GRASP.

\section{Conclusion}

We introduced an unsupervised motion-compensated reconstruction scheme for free‐breathing pulmonary MRI from highly undersampled measurements. The motion fields for each time frame were assumed to be generated by a deep CNN generator, driven by low-dimensional latent vectors. Each image in the time series was assumed to be a deformed version of an image template using the above motion fields. The parameters of the generator, latent vectors, and the image template were jointly estimated from the k-t space data. The comparison of the scheme with the state-of-the-art motion-resolved reconstruction demonstrates the preliminary utility of the proposed approach for motion-compensated free‐breathing pulmonary MRI reconstruction.

\section{Compliance with Ethical Standards}

This research study was conducted using human subject data. Approval was granted by the Ethics Committee of the institute where the data were acquired.

\bibliographystyle{IEEEbib}
\bibliography{refs}

\begin{thebibliography}{1}

\bibitem{feng2016xd}
Li~Feng, Leon Axel, Hersh Chandarana, Kai~Tobias Block, Daniel~K Sodickson, and
  Ricardo Otazo,
\newblock ``Xd-grasp: golden-angle radial mri with reconstruction of extra
  motion-state dimensions using compressed sensing,''
\newblock {\em Magnetic resonance in medicine}, vol. 75, no. 2, pp. 775--788,
  2016.

\bibitem{nakarmi2017kernel}
Ukash Nakarmi, Yanhua Wang, Jingyuan Lyu, Dong Liang, and Leslie Ying,
\newblock ``A kernel-based low-rank (klr) model for low-dimensional manifold
  recovery in highly accelerated dynamic mri,''
\newblock {\em IEEE transactions on medical imaging}, vol. 36, no. 11, pp.
  2297--2307, 2017.

\bibitem{ahmed2020free}
Abdul~Haseeb Ahmed, Ruixi Zhou, Yang Yang, Prashant Nagpal, Michael Salerno,
  and Mathews Jacob,
\newblock ``Free-breathing and ungated dynamic mri using navigator-less spiral
  storm,''
\newblock {\em IEEE Transactions on Medical Imaging}, vol. 39, no. 12, pp.
  3933--3943, 2020.

\bibitem{zou2021dynamic}
Qing Zou, Abdul~Haseeb Ahmed, Prashant Nagpal, Stanley Kruger, and Mathews
  Jacob,
\newblock ``Dynamic imaging using a deep generative storm (gen-storm) model,''
\newblock {\em IEEE Transactions on Medical Imaging}, 2021.

\bibitem{yoo2021time}
Jaejun Yoo, Kyong~Hwan Jin, Harshit Gupta, Jerome Yerly, Matthias Stuber, and
  Michael Unser,
\newblock ``Time-dependent deep image prior for dynamic mri,''
\newblock {\em IEEE Transactions on Medical Imaging}, 2021.

\bibitem{huttinga2021nonrigid}
Niek~RF Huttinga, Tom Bruijnen, Cornelis~AT van~den Berg, and Alessandro
  Sbrizzi,
\newblock ``Nonrigid 3d motion estimation at high temporal resolution from
  prospectively undersampled k-space data using low-rank mr-motus,''
\newblock {\em Magnetic resonance in medicine}, vol. 85, no. 4, pp. 2309--2326,
  2021.

\bibitem{zhu2020iterative}
Xucheng Zhu, Marilynn Chan, Michael Lustig, Kevin~M Johnson, and Peder~EZ
  Larson,
\newblock ``Iterative motion-compensation reconstruction ultra-short te (imoco
  ute) for high-resolution free-breathing pulmonary mri,''
\newblock {\em Magnetic resonance in medicine}, vol. 83, no. 4, pp. 1208--1221,
  2020.

\bibitem{johnson2013optimized}
Kevin~M Johnson, Sean~B Fain, Mark~L Schiebler, and Scott Nagle,
\newblock ``Optimized 3d ultrashort echo time pulmonary mri,''
\newblock {\em Magnetic resonance in medicine}, vol. 70, no. 5, pp. 1241--1250,
  2013.

\end{thebibliography}
\vspace{-1em}
\end{document}